\documentclass[12pt,preprint]{aastex}
\shorttitle{Surface density jumps as planet traps}
\shortauthors{F.\ S.\ Masset, A. Morbidelli, A. Crida \& J. Ferreira}
\begin{document}
\title{Disk surface density transitions as protoplanet traps}
\author{F.\ S.\ Masset\altaffilmark{1,2}} \affil{SAp, Orme des
Merisiers, CE-Saclay, 91191 Gif/Yvette Cedex, France}
\altaffiltext{1}{Also at IA-UNAM, Ciudad Universitaria, Apartado
Postal 70-264, Mexico DF 04510, Mexico} \email{fmasset@cea.fr}
\author{A. Morbidelli, A. Crida} \affil{Laboratoire Cassiop\'ee -
CNRS/UMR 6202, Observatoire de la C\^ote d'Azur, BP 4229, 06304 Nice
Cedex 4, France} 
\email{morby@obs-nice.fr, crida@obs-nice.fr}
\and 
\author{J. Ferreira} \affil{Laboratoire
d'Astrophysique de Grenoble, 414 rue de la piscine, BP 53, 38041
Grenoble Cedex 9, France} 
\email{Jonathan.Ferreira@obs.ujf-grenoble.fr}
\altaffiltext{2}{Send offprint requests to fmasset@cea.fr}
\begin{abstract}
  The tidal  torque exerted  by a protoplanetary  disk with  power law
  surface   density  and   temperature  profiles   onto   an  embedded
  protoplanetary embryo is generally a negative quantity that leads to
  the  embryo inwards  migration. Here  we investigate  how  the tidal
  torque balance  is affected at  a disk surface density  radial jump.
  The jump has two consequences :
  \begin{itemize}
  \item it affects the  differential Lindblad torque. In particular if
  the  disk  is merely  empty  on  the  inner side,  the  differential
  Lindblad torque almost amounts  to the large negative outer Lindblad
  torque.
  \item It  affects the  corotation torque, which  is a  quantity very
  sensitive to  the local  gradient of the  disk surface  density.  In
  particular if  the disk is  depleted on the  inside and if  the jump
  occurs radially  over a  few pressure scale-heights,  the corotation
  torque  is  a  positive quantity  that  is  much  larger than  in  a
  power-law disk.
  \end{itemize}
  We show  by means  of customized numerical  simulations of  low mass
  planets embedded  in protoplanetary  nebulae with a  surface density
  jump that the second effect is dominant, that is that the corotation
  torque  largely dominates  the differential  Lindblad torque  on the
  edge  of a central  depletion, even  a shallow  one. Namely,  a disk
  surface density  jump of about  $50$~\% over $3-5$  disk thicknesses
  suffices to cancel  out the total torque. {As a consequence the
  type I  migration of  low mass objects  reaching the jump  should be
  halted, and all these objects  should be trapped there provided some
  amount  of  dissipation  is  present  in the  disk  to  prevent  the
  corotation  torque  saturation.    As  dissipation  is  provided  by
  turbulence, which induces a jitter of the planet semi-major axis, we
  investigate  under which conditions  the trapping  process overcomes
  the trend  of turbulence to  induce stochastic migration  across the
  disk. We  show that  a cavity  with a large  outer to  inner surface
  density ratio efficiently traps embryos from $1$ to $15$~$M_\oplus$,
  at any radius up to $5$~AU  from the central object, in a disk which
  has same  surface density profile  as the Minimum Mass  Solar Nebula
  (MMSN).  Shallow surface density  transitions require light disks to
  efficiently trap embryos. In the case of the MMSN, this could happen
  in the very  central parts ($r < 0.03$~AU).}  We  discuss where in a
  protoplanetary  disk   one  can   expect  {a   surface  density
  jump}. This  effect could  constitute a solution  to the  well known
  problem that  the build up of  the first protogiant solid  core in a
  disk takes much longer than its type I migration towards the central
  object.
\end{abstract}
\section{Introduction\label{sec:intro}}
The migration of low  mass protoplanets ($M_p<15\;M_\oplus$) under the
action of disk tides is long known  to be a fast process in disks with
power  law  surface  density  profiles \citep{w97,tanaka}.   The  fast
migration  timescale estimates  of  these objects  even constitutes  a
bottleneck for the core accretion scenario, which implies a slow build
up  of  a  solid core  until  it  reaches  the mass  threshold  ($\sim
15\;M_\oplus$) above  which rapid  gas accretion begins.   Indeed, the
solid core build up  time is $10^{6-7}$~yrs \citep{petal96}, while the
migration  timescale of  a $M_p=1\;M_\oplus$  planet  is $O(10^5)$~yrs
\citep{w97,tanaka} and  scales inversely proportionally  to the planet
mass. The  existence of gaseous  giant planets at large  distances ($a
\sim  0.1-10$~AU)  from their  central  star  therefore constitutes  a
puzzle. Recent work by \cite{alibert05}  has shown that the core build
up time scale  can be lowered by taking  migration into account, which
prevents  the  depletion of  the  core  feeding  zone. However,  these
authors  find that  the most  up  to date  type~I migration  timescale
estimate, which includes three dimensional effects and the co-rotation
torque \citep{tanaka}, still needs to  be lowered by a factor $10-100$
in  order to  allow for  the solid  core survival.   The  total torque
exerted by the  disk onto the planet can be split  into two parts: the
differential Lindblad  torque, that corresponds  to the torque  of the
spiral wake  that the planet excites  in the disk,  and the corotation
torque,  exerted  by the  material  located  in  the planet  coorbital
region.  The role  of the corotation torque has  often been overlooked
in migration rate estimates. The two  main reasons for that is that it
is harder to evaluate than  the differential Lindblad torque, and that
it saturates (i.e. tends to  zero) in the absence of dissipation.  The
corotation torque scales with the radial gradient of $\Sigma/B$, where
$\Sigma$  is the disk  surface density  and $B$  is the  second Oort's
constant,  or half  the disk  flow vorticity  vertical component  in a
non-rotating  frame.   This  scaling  makes the  corotation  torque  a
quantity  very  sensitive to  local  variations  of  the disk  surface
density or rotation  profile. Here we investigate the  behavior of the
total (Lindblad + corotation) tidal  torque exerted on a planet in the
vicinity of a  surface density radial jump, in  order to investigate a
suggestion  by \citet{m02}  that localized,  positive  surface density
jumps  may be  able to  halt migration.   We assume  that  the surface
density  transition  occurs on  a  length  scale  $\lambda$ of  a  few
pressure scale heights $H$.  We consider the case in which the surface
density is  larger on  the outside  of the transition,  but we  do not
limit ourselves to the case where the surface density on the inside is
negligible compared  to its value  on the outer  side.  The case  of a
virtually  empty  central  cavity  has already  been  contemplated  by
\cite{KL02}  in  the  context,  different  of ours,  of  giant  planet
migration.   They conclude that  giant planet  migration is  halted or
considerably slowed down  once the planet is inside  the cavity and in
$2:1$ resonance  with the  cavity edge, as  beyond this  resonance the
disk torque onto a planet on a circular orbit becomes negligible.

In section~\ref{sec:analytic}  we provide simple  analytical estimates
of  the   Lindblad  and  corotation  torques  at   a  surface  density
transition.  We show  that the corotation torque, which  is a positive
quantity  there, is  likely to  overcome  the Lindblad  torque if  the
transition  is localized enough,  i.e.  $\lambda  < $  a few  $H$.  In
section~\ref{sec:setup} we  describe the numerical setup  that we used
to   check  this   prediction   with  a   numerical  hydro-code.    In
section~\ref{sec:num},  we  present   the  results  of  our  numerical
simulations  which indeed  exhibit for  a wide  range of  parameters a
fixed point at  the transition, i.e. a point  where the corotation and
Lindblad  torques  cancel each  other  and  where planetary  migration
stops.   {We  also discuss  in  this  section  the issue  of  the
saturation  of the  corotation torque  and the  need of  turbulence to
prevent it, and  the conditions under which turbulence  is able or not
to  unlock  a  planet  from   the  transition.}  We  then  discuss  in
section~\ref{sec:discuss} where  in protoplanetary disks  such surface
density transitions  can be  found, and what  are the  consequences of
these planet traps on giant planet formation.

\section{A simple analytic estimate}
\label{sec:analytic}
A protoplanet embedded in a gaseous protoplanetary disk excites in the
latter  a  one-armed spiral  wake~\citep{ol02},  as  a  result of  the
constructive  interference  of propagative  density  waves excited  at
Lindblad resonances with the planet.  This wake exerts a torque on the
planet,  which  can  be  decomposed  into the  outer  Lindblad  torque
($\Gamma_o$), which is negative and  that is exerted by the outer arm,
and the  inner Lindblad  torque ($\Gamma_i $),  which is  positive and
that is  exerted by the  inner arm.  These  two torques do  not cancel
out.    The  residue   $\Delta\Gamma=\Gamma_o+\Gamma_i$,   called  the
differential Lindblad torque, is negative~\citep{w86}, thereby leading
to  inward  migration. If  one  calls  one-sided  Lindblad torque  the
arithmetic mean of the absolute values of the outer and inner Lindblad
torques:

\begin{equation}
  \Gamma_{LR}=\frac{\Gamma_i-\Gamma_o}{2},
\end{equation}

then the  differential Lindblad  torque is a  fraction of  this torque
which  scales with the  disk aspect  ratio $h=H/r$,  where $H$  is the
pressure  scale height  or  disk  thickness and  $r$  the radius.   In
particular,  for  a  disk  with  uniform surface  density  and  aspect
ratio~\citep{w97}:

\begin{equation}
  \Delta\Gamma\approx-8h\Gamma_{LR}.
\end{equation}

As  noted  by  \cite{w97},   in  a  nebula  with  $h=O(10^{-1})$,  the
differential Lindblad  torque is a  sizable fraction of  the one-sided
Lindblad torque. This is of  some importance for our concern: denoting
$\Delta\Gamma'$ the Lindblad  torque on an empty cavity  edge (it then
amounts to  the outer Lindblad  torque, i.e. $\Delta\Gamma'=\Gamma_o$)
and $\Delta\Gamma$ the  Lindblad torque further out in  the disk where
we assume the surface density and aspect ratio profiles to be flat, we
have:
\begin{equation}
\label{eqn:lindbladratio}
\Delta\Gamma'\approx \Delta\Gamma/(8h),
\end{equation}

which means that the Lindblad torque on a cavity edge is at most $2-3$
times larger than further out in the disk, for $0.04<h<0.06$.

In addition  to the  wake torque, the  material in the  orbit vicinity
exerts  on the planet  the so-called  corotation torque.   This torque
scales   with   the   gradient   of   the   specific   vorticity   (or
vortensity)~\citep{gt79,wlpi91,wlpi92}:
\begin{equation}
\label{eqn:crscaling}
\Gamma_C\propto\Sigma\cdot\frac{d\log(\Sigma/B)}{d\log r}.
\end{equation}
This torque  comes from the  exchange of angular momentum  between the
planet and  fluid elements  that perform U-turns  at the end  of their
horseshoe streamlines  \citep{wlpi91,m01,m02}.  It has  been estimated
that   the   co-rotation   torque   amounts   to   a   few   tens   of
percent~\citep{kp93} of  the differential  Lindblad torque for  a disk
with  a power-law  surface  density profile.   \cite{tanaka} give  the
following  estimate for  a three  dimensional disk  with  flat surface
density and temperature profiles:
\begin{equation}
\label{eqn:crlindbladratio}
\Gamma_C \approx -(1/2)\Delta\Gamma
\end{equation}
Note  that the  corotation torque  cancels out  for a  surface density
profile $\Sigma\propto r^{-3/2}$, while  it is a positive quantity for
a shallower surface density profile.

We now estimate the ratio of the corotation torque at a cavity edge
and further out in the disk (assuming a flat surface density
profile). We assume that the surface density inside of the transition
is $\Sigma_i$, while it is $\Sigma_o>\Sigma_i$ on the outside. The
transition from $\Sigma_i$ to $\Sigma_o$ occurs on a length scale
$\lambda\ll r$.  In the outer disk where $\Sigma \equiv \Sigma_o$, we have:
\begin{equation}
\frac{d\log(\Sigma/B)}{d\log r}=3/2,
\end{equation}
while at the transition, $d\log\Sigma/d\log r$ is a sharply peaked function with a
maximum that scales as:
\begin{equation}
\left.\frac{d\log\Sigma}{d\log r}\right|_{\rm max}\sim\frac r\lambda\log\left(\frac{\Sigma_o}{\Sigma_i}\right).
\end{equation}
Similarly, retaining only the terms that vary most rapidly with~$r$ at the transition:
\begin{equation}
\frac{d\log B}{d\log r}\simeq \frac {2r^2}{\Omega}\frac{d^2\Omega}{dr^2},
\end{equation}
where $\Omega$ is the angular velocity. The last factor of the above
equation can also be expressed, using the disk rotational equilibrium
and again keeping only the terms that vary most rapidly with $r$, as:
\begin{equation}
\frac{d^2\Omega}{dr^2}\simeq-\frac{\Omega h^2r}{2}\frac{d^3\log\Sigma}{dr^3}.
\end{equation}
The last  factor of  the R.H.S. is  a function  of $r$ that  scales as
$(H^2r/\lambda^3)\log(\Sigma_o/\Sigma_i)$,   and  that   has,   for  a
sufficiently   smooth,  monotonic   transition   from  $\Sigma_i$   to
$\Sigma_o$,  two  minima and  one  maximum,  the  exact locations  and
amplitudes  of  which depend  on  the  surface  density profile.   The
(inverse of) the specific vorticity logarithmic gradient is therefore:

\begin{equation}
\frac{d\log(\Sigma/B)}{d\log r}=\frac r\lambda\log
\left(\frac{\Sigma_o}{\Sigma_i}\right)
\left(a+b\frac{H^2}{\lambda^2}\right),
\end{equation}
where $a$ and  $b$ are numerical functions of $r$  of order unity that
depend  on  the  exact  shape  of the  surface  density  profile.   In
particular,  one  sees  immediately   that  if  $\lambda\sim  H$,  the
contributions  of both  terms (the  surface density  gradient  and the
vorticity gradient) to the corotation torque are of similar magnitude.

The corotation torque at the transition $\Gamma_C'$ can therefore be
larger than the corotation torque $\Gamma_C$ further out in the disk
by a factor:
\begin{equation}
\label{eqn:crratio}
\frac{\Gamma_C'}{\Gamma_C}\approx\frac{2r}{3\lambda}C,
\end{equation}
where $C$  is a numerical  factor of order  unity that depends  on the
surface density profile.

Using  Eqs.~(\ref{eqn:lindbladratio}), (\ref{eqn:crlindbladratio}) and
(\ref{eqn:crratio}), we find that:
\begin{equation}
\frac{\Gamma_C'}{\Delta\Gamma'}\approx-\frac{8H}{3\lambda}C,
\end{equation}

which means  that the (positive) corotation torque  can counteract the
(negative) Lindblad torque at a  cavity, provided that the edge of the
latter be  narrow enough  ($\lambda < 8HC/3$).   We check by  means of
numerical  simulations  that it  is  indeed  possible  to halt  type~I
migration on the edge of a surface density transition.

\section{Numerical set up}
\label{sec:setup}
We  performed   a  large  number   of  two-dimensional  hydrodynamical
simulations   that    we   detail   below.    The    code   used   was
FARGO\footnote{See: {\tt http://www.star.qmul.ac.uk/$\sim$masset/fargo}}.
It is a staggered mesh code on a polar grid, with upwind transport and
a harmonic, second  order slope limiter \citep{vl77}.  It  also uses a
change of  rotating frame  on each ring  that enables one  to increase
significantly the time step \citep{m00,m00b}.

The surface  density transition  was set in  the following  manner. We
call $F$ the  target surface density ratio between  the inside and the
outside of the cavity: $F=\Sigma_o/\Sigma_i$. In order to realize this
ratio, we adopt a fixed profile of kinematic viscosity with an inverse
ratio: $\nu_o/\nu_i=F^{-1}$. If  we call $r_t$ the mean  radius of the
transition,   then   for   $r>r_t+\lambda/2$,  $\nu\equiv\nu_o$,   for
$r<r_t-\lambda/2$,          $\nu\equiv\nu_i$,          and         for
$r_t-\lambda/2<r<r_t+\lambda/2$, $\nu$  is an affine  function of $r$.
We  then relax  the surface  density profile  during a  preliminary 1D
calculation over $\sim 10^5$ orbits, so that the profile can safely be
considered as steady state in the 2D calculations. In our calculations
we use  non-reflecting inner and  outer boundary conditions,  while we
impose the disk surface density at the inner and outer boundary.

In all our calculations the disk aspect ratio is $h=0.05$, the disk is
not  self-gravitating and  it  has a  locally  isothermal equation  of
state.  The length unit is  $r_t$. The mesh inner and outer boundaries
are respectively at  $r=0.42$ and $r=2.1$.  The mass  unit is the mass
$M_*$ of the central object,  and the gravitational constant is $G=1$.
The time  unit is $(GM_*/r_t^3)^{-1/2}$.   Whenever we quote  a planet
mass in  Earth masses, we  assume the central  object to have  a solar
mass.

The mesh size in all our runs is $N_r=168$ and $N_\phi=450$, hence the
radial  resolution  is $10^{-2}$.   The  horseshoe  zone  width for  a
$15\;M_\oplus$ planet  is $\sim 7.8\cdot  10^{-2}$, so it is  $\sim 8$
zones wide.   The error  on the corotation  torque due to  this finite
width  therefore amounts  to  at most  $\sim  10$~\% [see  \cite{m02},
Fig. A.3].  We also made calculations with a $4\;M_\oplus$ mass planet
for which the  horseshoe zone is only $4$  zones wide. This resolution
may seem  low but the corotation  torque is very well  captured at low
resolution  for small mass  planets which  hardly perturb  the surface
density  profile~[see  discussion   in  \cite{m02},  Appendix~A].   

We  contemplated  the fashionable  issue  of  whether  the Roche  lobe
content must be  included or not when integrating  the torque over the
disk material \citep{abl05,mp03}.  We  found that taking that material
into account or  not makes essentially no difference  in our case.  We
make the following technical comments related to this issue:
\begin{itemize}
\item \citet{abl05} have envisaged the  case of Jupiter or Saturn mass
planets, for which there is a significant Roche lobe with radius $\sim
a(M_p/3M_*)^{1/3}$  ($a$ being  the planet  semi-major axis).  For the
mass range  studied here ($M_p<15\;M_\oplus$), the Roche  lobe size is
much smaller than  this simple estimate. In particular,  in the linear
limit, the flow does  not exhibit any circumplanetary libration, while
the  two ends  of  the  horseshoe region  reconnect  through a  single
stagnation  point located  on  the  orbit (Masset  et  al., in  prep.;
Tanaka, private  communication), that is  offset from the planet  as a
result of the gaseous disk slight sub-Keplerianity. The issue of Roche
lobe sampling in that case is therefore irrelevant.
\item \citet{abl05}  have found that  resolution may alter  the torque
exerted  on the  planet when  the Roche  lobe material  is  taken into
account.  This behavior,  which appears  at very  high  resolution, is
related to  the build up of  a gaseous envelope  around the point-like
planet,  while the  mass of  this  envelope heavily  depends upon  the
equation of  state, which is poorly  known. The envelope  mass may be
considerable for  the giant protoplanets  considered by \citet{abl05},
as it  may be several  times the planet  mass, but it is  presumably a
much smaller fraction of the planet mass for the planet masses that we
consider [$M_p<15\;M_\oplus$; see also \citet{alibert05}].

\end{itemize}

In 2D  simulations one has  to introduce a potential  smoothing length
$\epsilon$  meant  to mimic  the  disk  vertical  extent.  The  planet
potential expression is then $-GM_p/(r^2+\epsilon^2)^{1/2}$, where $G$
is the gravitational constant and  $r$ the distance to the planet.  In
our calculations we use  a potential smoothing length $\epsilon=0.7H$.
This  is  quite  a  large  value, which  leads  to  underestimate  the
corotation torque \citep{m02}, as  it lowers the horseshoe zone width,
while the  corotation torque  is proportional to  the fourth  power of
this width.   Our calculations are  therefore conservative, and  it is
likely that 3D  calculations would yield a torque  cancellation for an
even wider set of disk parameters than what we found.

\section{Results of numerical simulations}
\label{sec:num}
\subsection{Torque sampling on a cavity edge}
\label{sec:numcavit}
We  consider in  this section  a disk  with a  target  surface density
ratio between  the inside  and the outside  of the transition  set to
$F=7$. As this value is large compared to one, we call the inside part
of  the  transition  a  cavity.  We  performed  two  sequences  of  75
calculations in which a planet was  set on a fixed circular orbit with
radius $a_i=0.65+10^{-2}i$ ($i=0\ldots 74$). For these calculations we
report as a  function of $a_i$ the total torque  exerted on the planet
[averaged over  $100$~orbits, in order  for the libration  flow inside
the co-orbital region to reach a steady state~\citep{m02}].

\begin{figure}
\plottwo{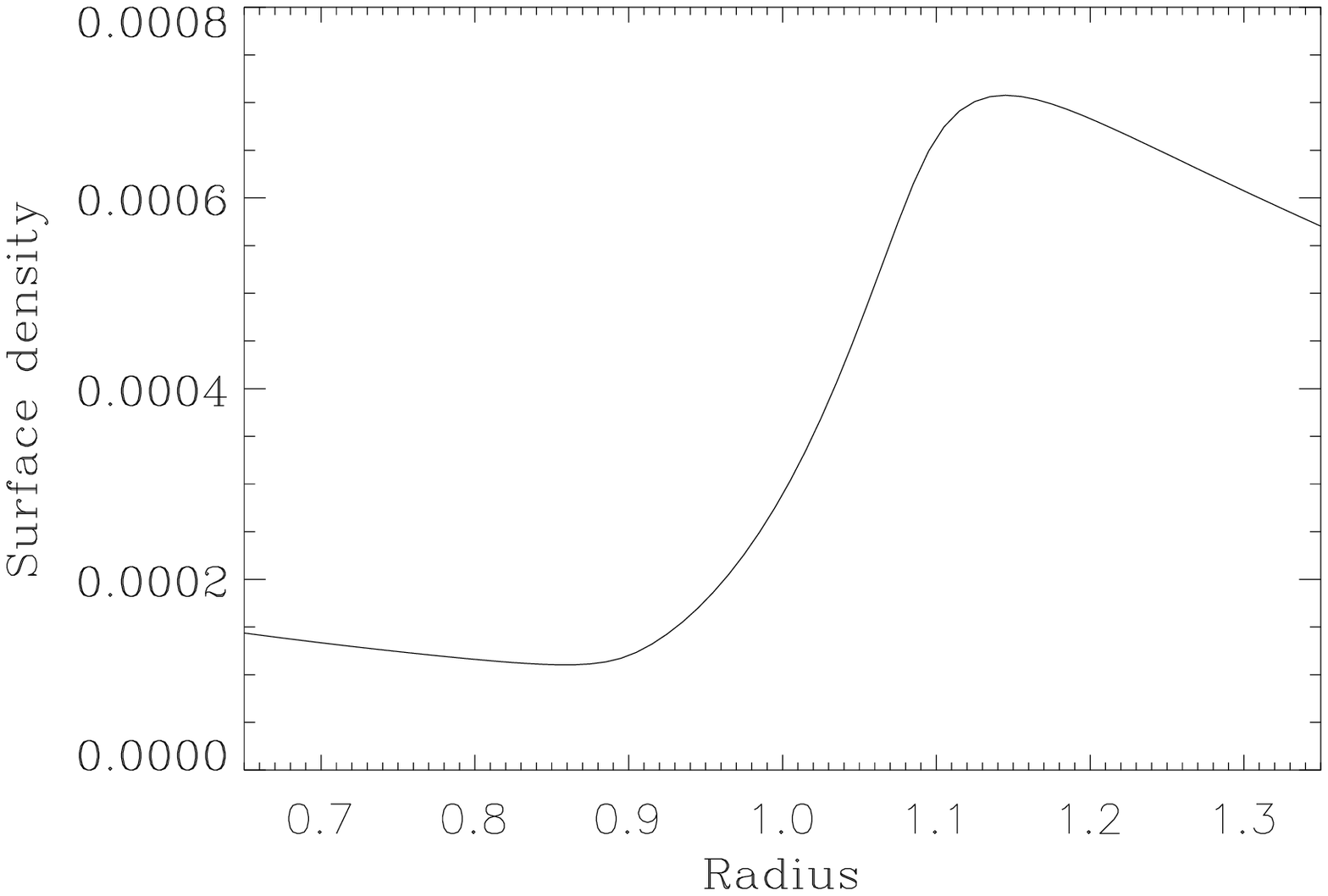}{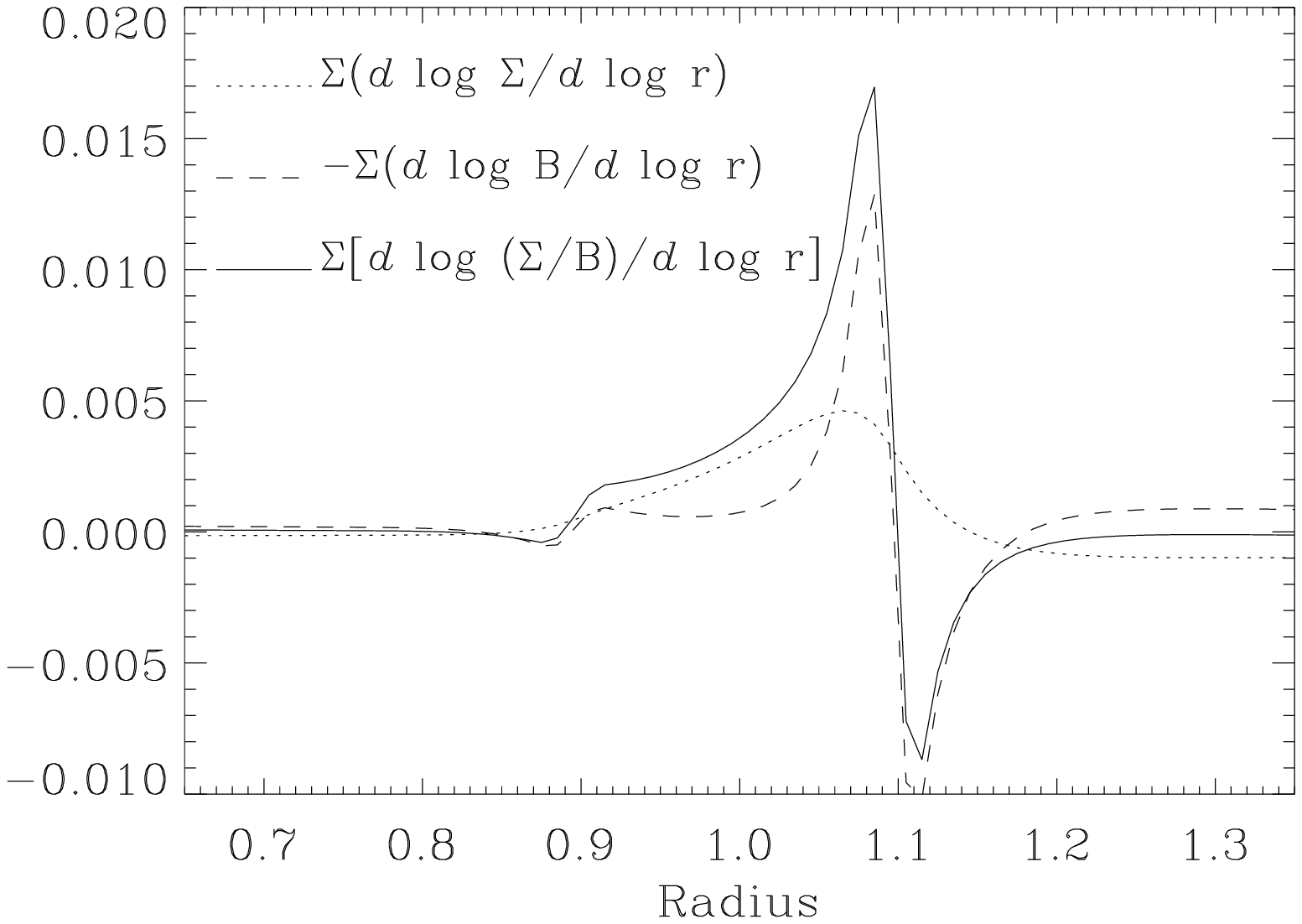}
\caption{\label{fig:cavitprop}Left:  surface density  profile  for the
runs  of section~\ref{sec:numcavit}.  The  surface density  transition
occurs  over   $\sim  4H$,   between  $r\approx  0.9$   and  $r\approx
1.1$.  Right:  Corresponding  profiles  of the  different  logarithmic
derivatives  involved  in the  expression  of  the corotation  torque,
weighted  by   the  surface  density.   The  dotted   line  shows  the
logarithmic derivative  of the surface density, the  dashed line shows
the opposite of the logarithmic  derivative of the flow vorticity, and
the solid line shows the  sum of these two quantities.  It corresponds
to the R.H.S.  of Eq.~(\ref{eqn:crscaling}).}
\end{figure}

The left  plot of  Fig.~\ref{fig:cavitprop} shows the  relaxed surface
density profile after $10^5$ orbits of the preliminary 1D calculation,
and the  right plot  shows the quantities  involved in  the corotation
torque  scaling.  The  right panel  of  Fig.~\ref{fig:cavitprop} shows
that the vortensity gradient peak is only partially due to the surface
density gradient,  since the  gradient of the  flow vorticity  plays a
role     of     comparable     importance,     as     suggested     in
section~\ref{sec:analytic}.

\begin{figure}
\plottwo{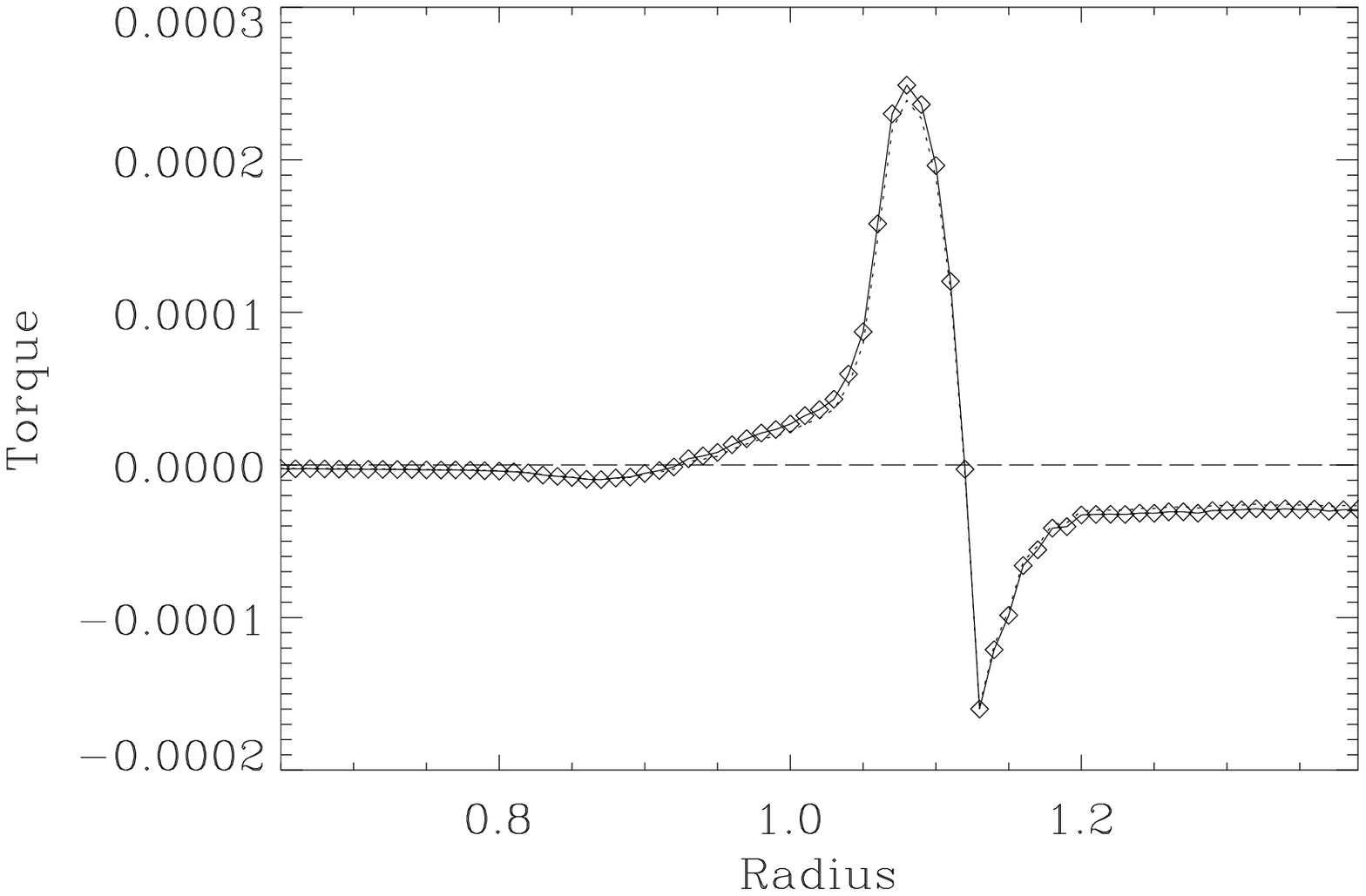}{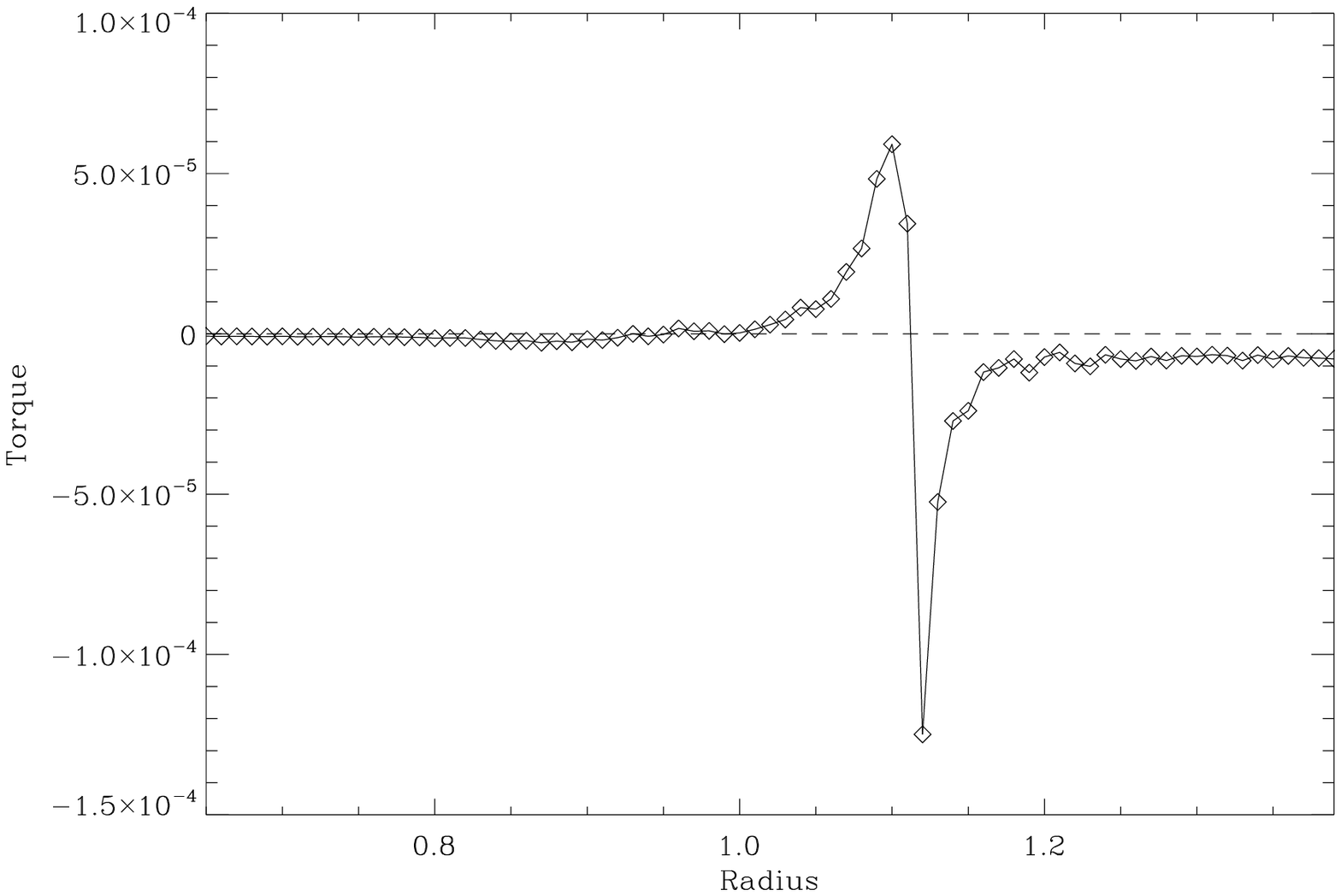}
\caption{\label{fig:tqnumcavit}Left: Specific torque acting upon a
$15\;M_\oplus$ planet on a fixed circular orbit, as a function of
its orbital radius. Right: same thing for a $4\;M_\oplus$ planet.}
\end{figure}

The total torque measured from numerical simulations is presented on
Fig.~\ref{fig:tqnumcavit}. Both plots, which correspond to two different
planet masses, exhibit similar features:
\begin{itemize}
\item  On the  left side  (for  $r<0.8$) and  on the  right side  (for
$r>1.2$), the torque profile is flat and negative. This corresponds to
the  total negative  torque  acting on  a  planet embedded  in a  flat
surface density disk.  Since the  surface density is $F=7$ times lower
on the inside, the torque value is much smaller on that side.
\item In  the range  $0.8<r<1.2$, the torque  as a function  of radius
exhibits  a shape  very similar  to the  vortensity gradient  shown on
Fig.~\ref{fig:cavitprop}. In particular the torque has a maximum value
near $r=1.08-1.10$ and a minimum value near $r=1.12-1.13$.
\end{itemize}
This  last  observation unambiguously  indicates  that the  corotation
torque  largely dominates  the  total torque  at  the surface  density
transition, as  expected from  the analytic estimate.   In particular,
there are two positions in the disk where the total torque cancels out
(and therefore  where the migration is halted):  one at $r\approx0.92$
and the  other one  at $r\approx1.11-1.12$. The  first of  these fixed
points is  unstable : moving the  planet away from it  yields a torque
that moves it further away,  while the second one (i.e.  the outermost
one) is stable: moving the planet to a higher radius yields a negative
torque that tends to bring the planet back to its former position, and
vice-versa for  a motion  towards smaller radii.   The fixed  point at
$r=1.12$ should  therefore trap all type~I migrating  embryos. We note
in  passing that  the torque  as a  function of  radius does  not have
exactly the  same shape between  a $15\;M_\oplus$ and  a $4\;M_\oplus$
planet.   There are  two reasons  for  this: the  onset of  non-linear
effects between  $4\;M_\oplus$ and $15\;M_\oplus$,  and the inaccuracy
of the corotation torque estimate  for the $4\;M_\oplus$ case, for the
resolution  that we have  adopted. However,  we see  that even  in the
$4\;M_\oplus$  case, the  value of  the torque  maximum  ($\sim 6\cdot
10^{-5}$) is much larger than  the torque absolute value away from the
cavity (for $r>1.2$, this value is $\sim10^{-5}$). Therefore, even if
the error on the corotation torque amounts to $50$~\% [see \cite{m02}],
one still gets a torque cancellation and a stable fixed point.

\subsection{Halting migration at a cavity edge}
\begin{figure}
\plottwo{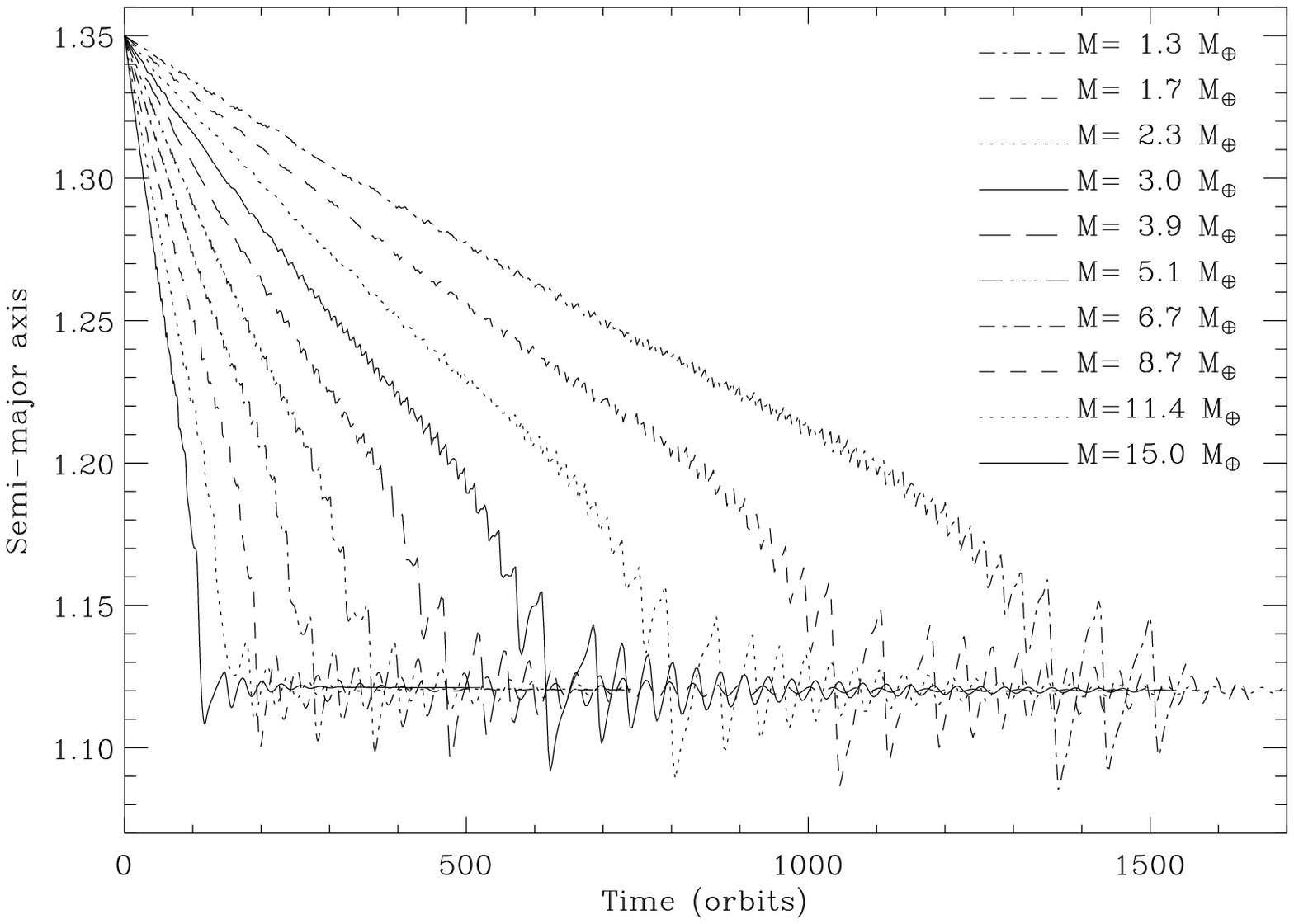}{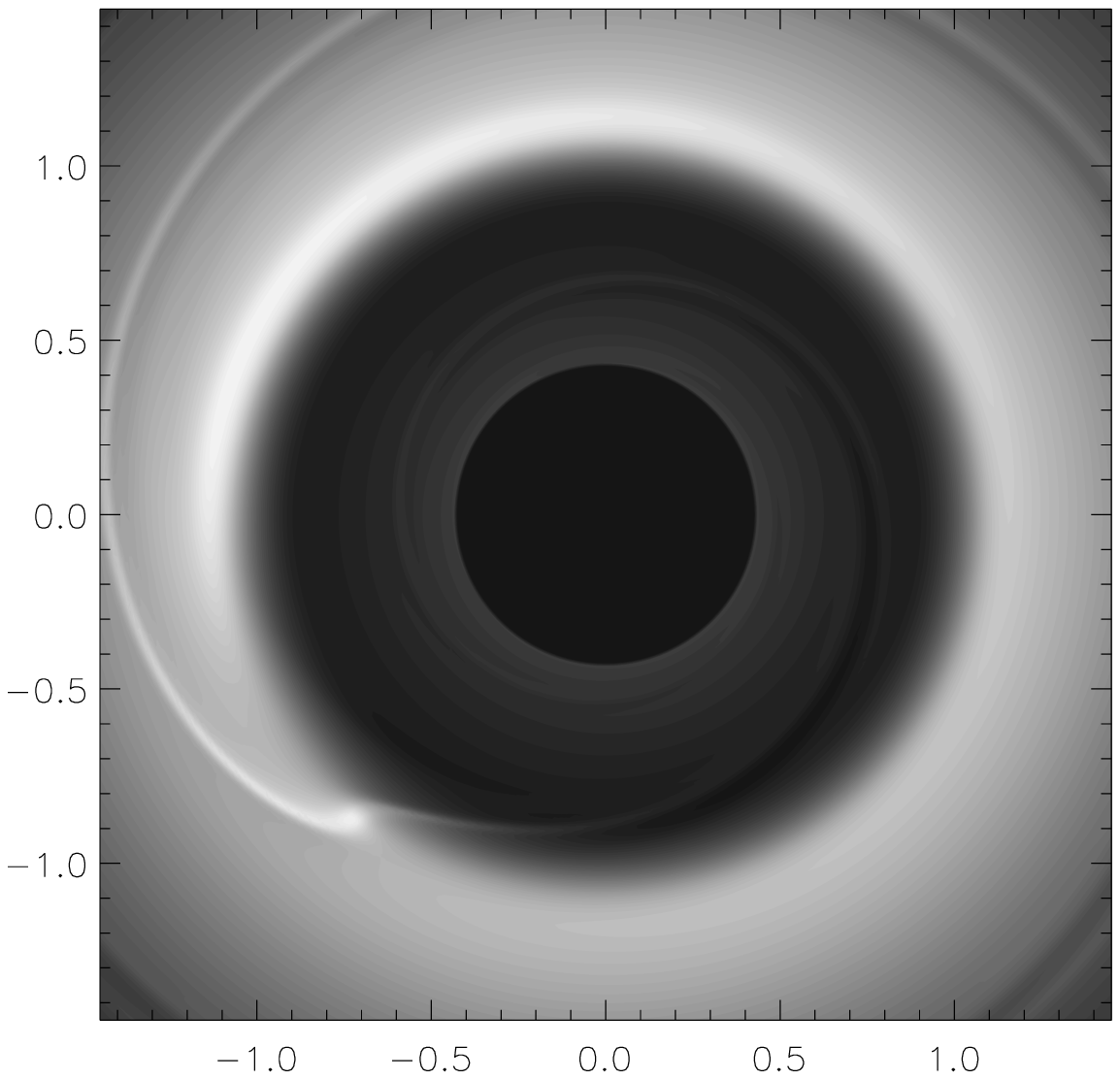}
\caption{\label{fig:freemig}Left plot: primary to planet separation
curve as a function of time, for the 10 different planet
masses. Right plot: snapshot of the surface density map with a
$15\;M_\oplus$ planet embedded at $t=480$~orbits.}
\end{figure}

We have performed a series  of ten calculations with planets of masses
$1.3\;M_\oplus$ to $15\;M_\oplus$ in geometric sequence, with the same
disk cavity as in the  previous section ($F=7$). In these calculations
the planets  were released at  $r=1.35$, i.e. beyond the  cavity edge,
and  then allowed  to  freely migrate  under  the action  of the  disk
torque.  The results are  presented at Fig.~\ref{fig:freemig}.  We see
how the  transition efficiently traps  all these objects at  the fixed
point  at $r=1.12$  identified at  the  previous section.  We note  in
passing that on the lower side of the mass range considered, the small
width of the  horseshoe region may lead to  large uncertainties on the
corotation torque. Nevertheless, the migration of all objects stops at
locations which are all very  close to $1.12$. This indicates that the
corotation  torque at  the  edge,  even misrepresented  by  a too  low
resolution, still strongly dominates the differential Lindblad torque.
For masses  lower than  a few  earth masses, the  Hill radius  is much
smaller than  the disk thickness and  the disk response  is linear. In
the  linear  regime,  the  corotation  to  Lindblad  torque  ratio  is
independent  of the  planet mass.   As we  have stressed  already, our
potential smoothing length is large and our corotation torque estimate
is conservative.  This indicates that any small mass in-falling embryo
will be trapped at the cavity edge that we modeled.

Disk profiles with an extremum  of the specific vorticity are prone to
a  linear  Rossby  wave  instability  that  eventually  leads  to  the
formation of vortices~\citep{li2000,li2001}.  This  is the case of the
transitions that  we have simulated. The  arc-shaped over-dense region
located    near    $x=-0.5,y=+0.5$    in    the    right    plot    of
Fig.~\ref{fig:freemig} corresponds  to such a  vortex.  It leads  to a
torque modulation over  the synodic period between the  planet and the
vortex,  that is  seen on  the migration  curves of  the left  plot of
Fig.~\ref{fig:freemig}.   As a planet  approaches the  transition, the
synodic period  increases, and  so does the  period of  the semi-major
axis modulations.

\subsection{Torque sampling at a shallow surface density jump}
\label{sec:numshallow}
\begin{figure}
\plottwo{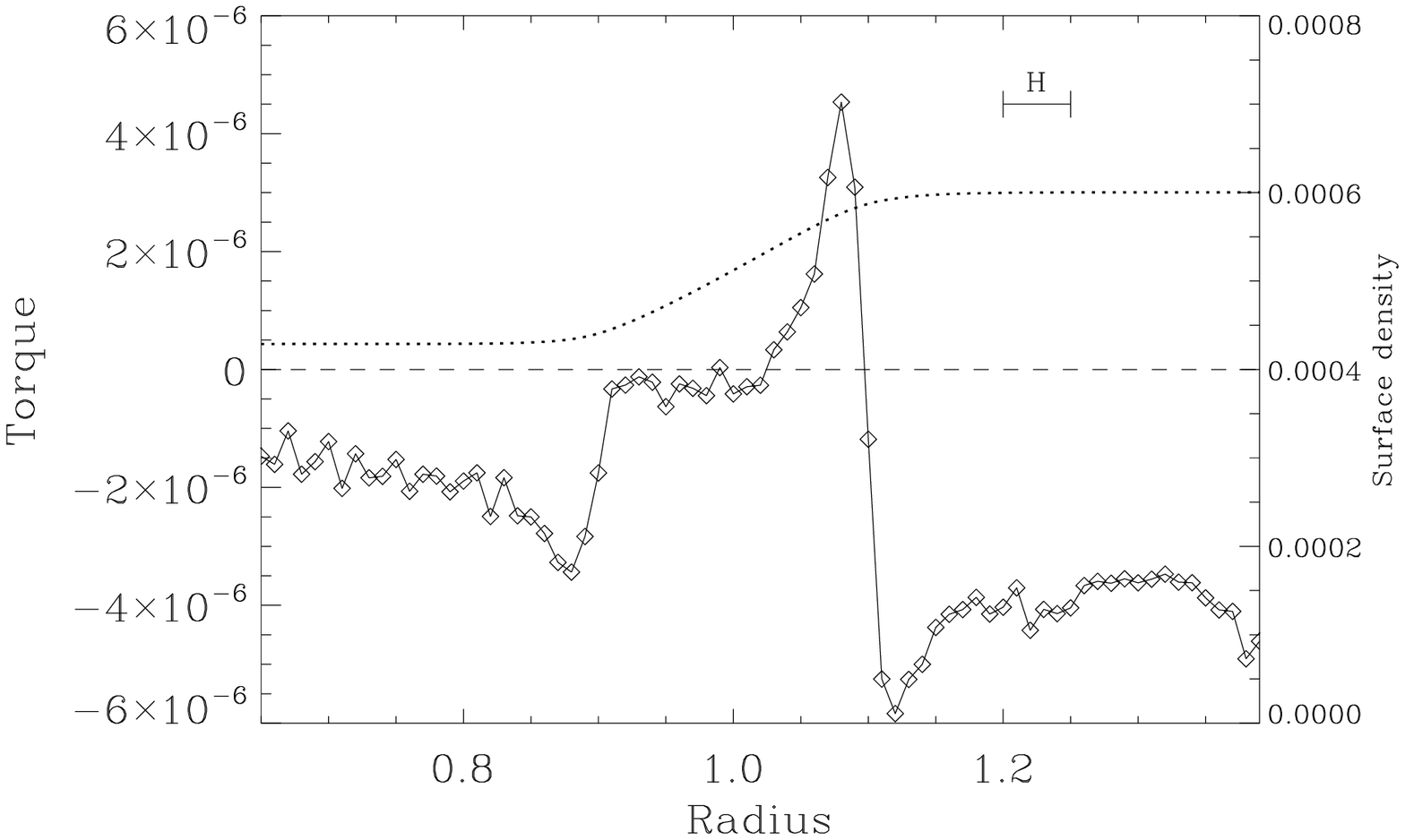}{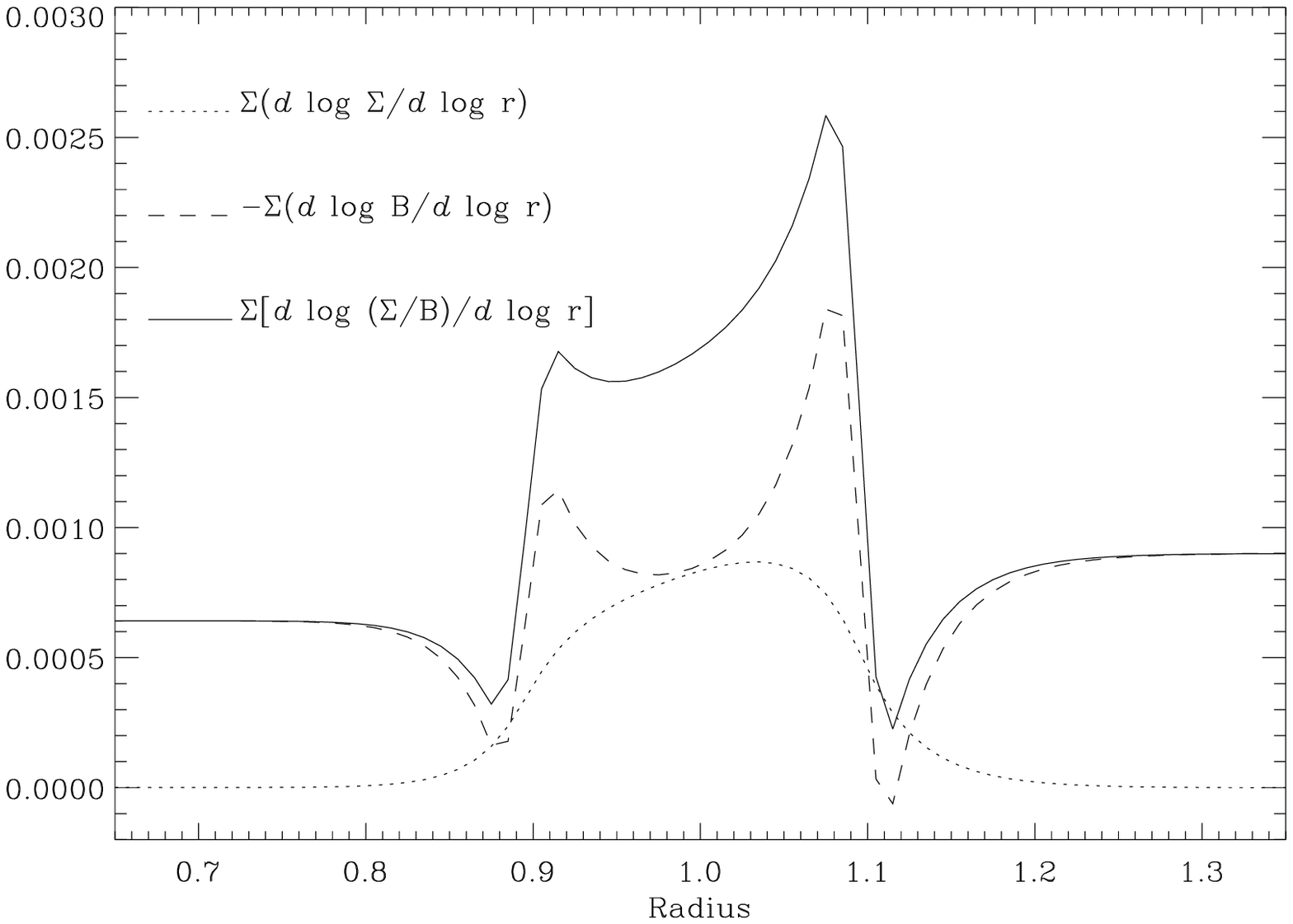}
\caption{\label{fig:cavitshallow}Left: surface density profile for the
runs of section~\ref{sec:numshallow} (dotted  line and right axis) and
total torque  exerted by the disk on  the planet as a  function of its
semi-major axis  (diamonds and left axis), for  a 4~$M_\oplus$ planet.
Right: Corresponding profiles of the different logarithmic derivatives
involved in the  expression of the corotation torque,  weighted by the
surface density, as in Fig.~\ref{fig:cavitprop}.}
\end{figure}

We  have  performed a  series  of  calculations  similar to  those  of
section~\ref{sec:numcavit},  except  that we  adopted  a much  smaller
target outer to  inner surface density ratio: $F=1.4$  instead of $7$.
The surface density jump is  therefore a very shallow one. The results
are  presented on  Fig.~\ref{fig:cavitshallow}. They  show  that again
there  is a fixed  point at  which migration  is halted,  at $r\approx
1.09$. We have  not searched further the limit  inner to outer surface
density  ratio that  leads  to migration  reversal,  as this  quantity
depends  on the  jump  radial size,  on  the profile  adopted for  the
kinematic  viscosity and on  the potential  smoothing length.  We just
stress that a factor $F=1+O(1)$ and  a jump size of $\sim 4H$ suffices
to halt migrating bodies.

\subsection{Driving of trapped planets}
\label{sec:trap}
\begin{figure}
\plotone{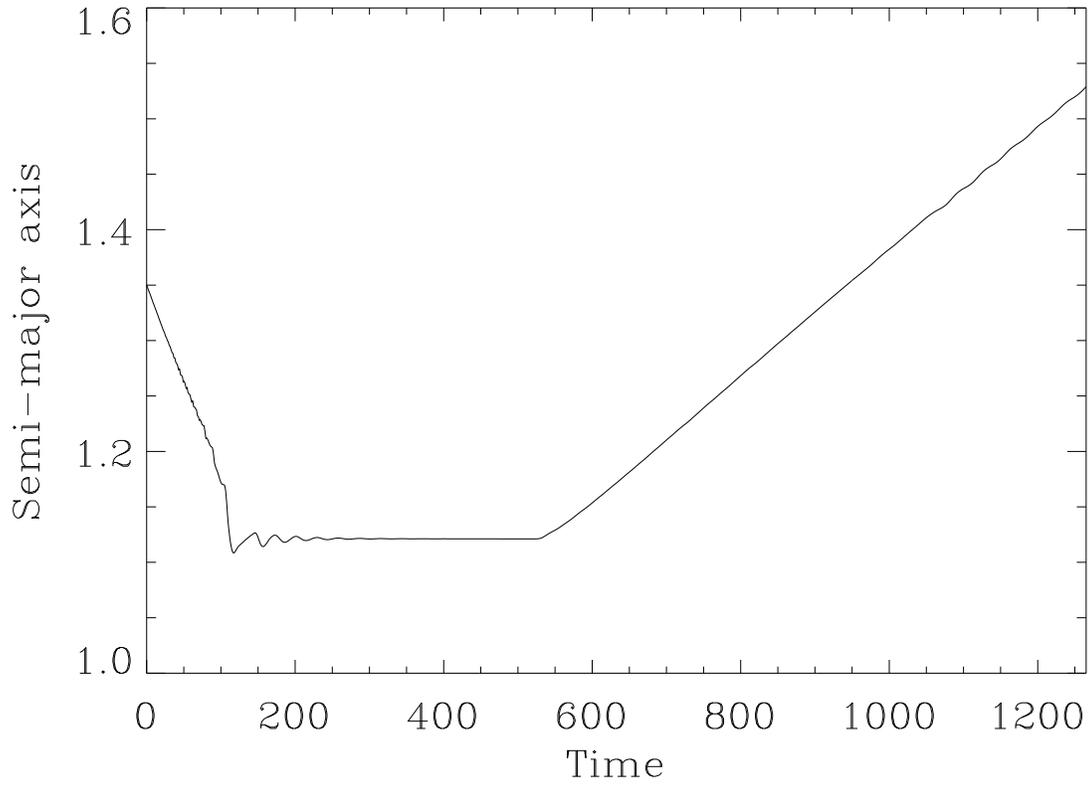}
\caption{\label{fig:trap}Semi-major axis  as a function of  time for a
planet trapped  in an  expanding cavity. The  cavity is fixed  for the
first $520$~orbits,  then is linearly expanding with  time. The planet
semi-major  axis  follows the  cavity  evolution  so  that the  planet
remains trapped  at the cavity edge.  At $t=0$ the  planet is released
beyond  the cavity,  at $r=1.35$.  During  the first  $100$ orbits  it
migrates towards the cavity where it gets trapped.}
\end{figure}

As can be seen on Fig.~\ref{fig:tqnumcavit} or~\ref{fig:cavitshallow},
the torque maximum value is  positive and large. This implies that, if
the cavity  radius evolves with  time, either inwards or  outwards, it
drives any  planet trapped on its  edge so that  its migration follows
the  edge  evolution. More  precisely,  a planet  will  be  held at  a
location where  the specific  torque acting upon  it endows it  with a
drift rate equal  to that of the cavity radius.  As the torque maximum
in absolute value is at least of  the order of the total torque in the
outer disk, this  means that the edge expansion  or contraction can be
as fast as the type~I drift rate of its trapped objects and still keep
these   objects   trapped.   We   illustrate   this   shepherding   on
Fig.~\ref{fig:trap}. We performed a  calculation in which an initially
fixed cavity such as the one considered at the previous sections (here
with  $F=10$ and  a  transition over  $4-5H$)  traps a  $15\;M_\oplus$
in-falling planet  and then  is endowed with  an expansion  rate $\dot
r_t=8\cdot  10^{-5}$.  This  is   achieved  by  imposing  a  kinematic
viscosity  profile  as  described  in section~\ref{sec:setup}  with  a
radius  $r_t$ that varies  linearly with  time. One  can see  that the
planet semi-major  axis follows the  cavity radius evolution,  so that
the planet remains trapped on the cavity edge.

\subsection{Dissipation and corotation torque saturation}
\subsubsection{Corotation torque saturation}
The corotation torque, in the absence of any angular momentum exchange
between the  horseshoe region and  the rest of  the disk, is  prone to
saturation:   its  value  tends   to  zero   after  a   few  libration
periods~\citep{wlpi91,wlpi92,m01,bk01,m02}. The disk turbulence, which
endows the disk  with an effective viscosity, can  help to prevent the
corotation torque saturation. Here we estimate the amount of viscosity
required to prevent saturation, as  a function of the planet mass. The
maximal  amount  of  saturation  for  which one  still  has  a  torque
cancellation  depends on  the planet  mass  and on  the detailed  disk
profiles. Here  we just  give an order  of magnitude of  the viscosity
needed to  prevent saturation by  assuming a $50$~\%  saturation. This
amount of saturation is reached  when the viscous timescale across the
horseshoe    region   is    equal   to    the    horseshoe   libration
timescale~\citep{m01}:
\begin{equation}
\label{eqn:sat}
\frac{x_s^2}{3\nu}=\frac{4\pi a}{(3/2)\Omega_px_s},
\end{equation}
where  $x_s$ is  the  horseshoe zone  half  width, $\nu$  is the  disk
effective  kinematic  viscosity,  $\Omega_p$  is  the  planet  orbital
frequency and $a$ its semi-major axis.  In order to express $x_s$ as a
function  of the  planet mass\footnote{The  estimate provided  here is
given for  a flat  surface density profile,  but the  horseshoe region
width should not  or very weakly depend on  this assumption.}, we note
that  in the  linear (hence  fully unsaturated)  limit  the corotation
torque can be expressed as~\citep{wlpi91,m01}:
\begin{equation}
\label{eqn:tq1}
\Gamma_C=\frac 98x_s^4\Omega_p^2\Sigma,
\end{equation}
while from \cite{tanaka} we have:
\begin{equation}
\label{eqn:tq2}
\Gamma_C=0.939h^{-2}q^2a^4\Omega_p^2\Sigma,
\end{equation}
where $q=\frac{M_p}{M_*}$.
Eqs.~(\ref{eqn:tq1}) and (\ref{eqn:tq2}) yield:
\begin{equation}
\label{eqn:xs}
x_s=0.96a\sqrt\frac qh.
\end{equation}
Eqs.~(\ref{eqn:sat})  and~(\ref{eqn:xs})  give  the minimum  effective
viscosity to prevent the corotation torque saturation:
\begin{equation}
\nu=0.035 \left(\frac qh\right)^{3/2}a^2\Omega_p,
\end{equation}
or, in terms of the $\alpha$ coefficient ($\nu=\alpha H^2\Omega_p$):
\begin{equation}
\alpha=0.035 q^{3/2}h^{-7/2}=6.5\cdot
10^{-6}\cdot\left(\frac{M_p}{1\;M_\oplus}\right)^{3/2}
\cdot\left(\frac{h}{0.05}\right)^{-7/2}
\end{equation}
The  value  of $\alpha$  required  to  prevent  the corotation  torque
saturation of an Earth mass  planet is therefore much smaller than the
one   inferred  from  disk   lifetimes  of   T  Tauri   stars,  namely
$\alpha=10^{-3}-10^{-2}$~\citep{hartmann01}.   Note   that  this  also
holds  for  large   mass  solid  cores  ($M_p=10-15\;M_\oplus$).   The
corotation torque acting on these  objects should therefore be a large
fraction of the unsaturated  estimate, so the analysis presented above
for the torque  cancellation is essentially valid.  We  also note that
the  planets   of  Figs.~\ref{fig:freemig}  and~\ref{fig:trap}  remain
trapped  over a  timescale much  longer than  the  horseshoe libration
timescale, so that their  corotation torque partial saturation is weak
enough. The disk effective viscosity  at the trap location is in these
calculations $5\cdot 10^{-6}$ (i.e.  $\alpha=2\cdot 10^{-3}$).

{We  comment  that  the  corotation torque  saturation  could  be
different in  a laminar disk  with kinematic viscosity~$\nu$ and  in a
turbulent  disk (we  discuss more  in detail  this topic  in  the next
section) with the same  effective viscosity, when the horseshoe region
width  is  smaller than  the  turbulence  length scale. The  saturation
should be easier to remove in  this last case, which suggests that our
saturation estimates is a conservative one.}

\subsubsection{Dissipation and turbulence}

\begin{figure}
\plotone{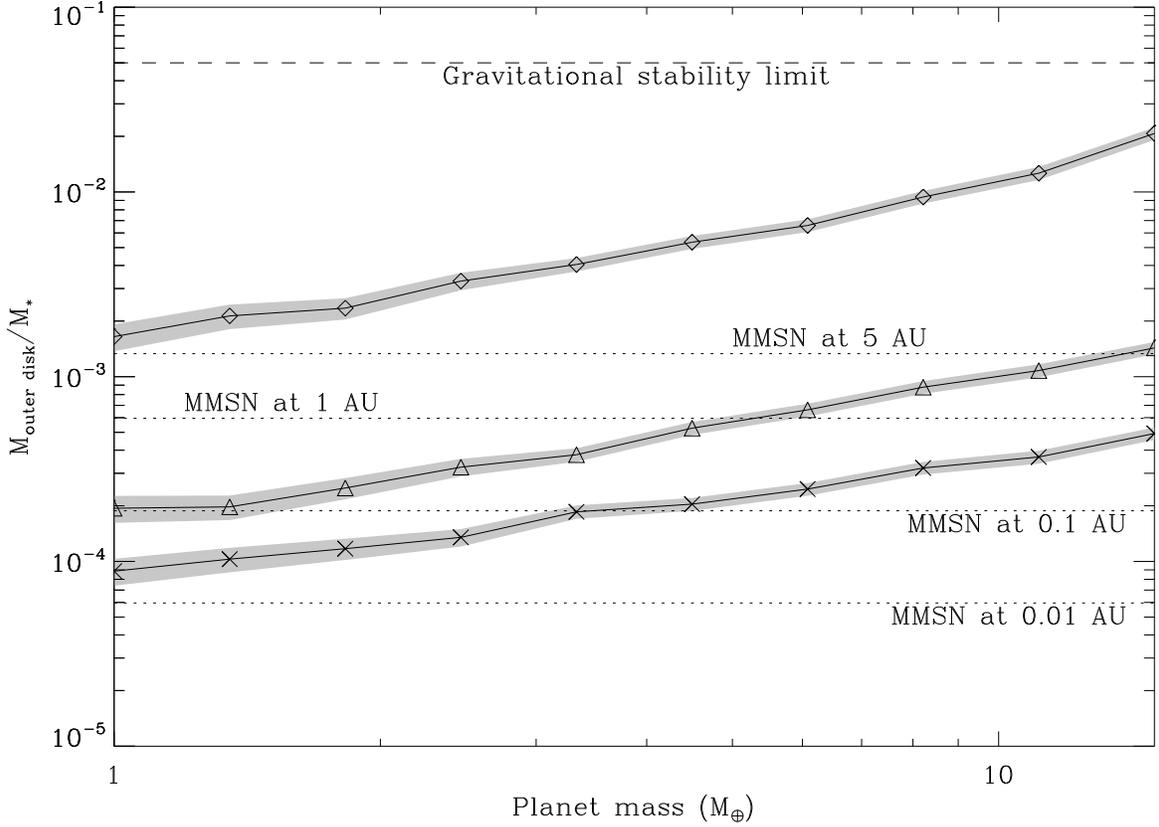}
\caption{\label{fig:trapturb} Critical disk mass  as a function of the
planet  mass  (from $1\;M_\oplus$  to  $15\;M_\oplus$), for  different
transition  profiles: the  $F=7$ cavity  of section~\ref{sec:numcavit}
(diamonds), the  $F=1.4$ shallow jump  of section~\ref{sec:numshallow}
(crosses).  The triangles represent  the trapping  limit for  a torque
curve with  a peak  maximum which  is the geometric  mean of  the peak
values of  the $F=7$  and $F=1.4$ transitions,  which we assume  to be
representative  of   a  transition  with   $F\simeq  1+\sqrt{0.4\times
6}=2.5$. The dotted lines show  the minimum mass solar nebula relative
mass (as defined by its local surface density) at different radii. The
thickness of  the gray line  indicates the results  standard deviation
found with different random seeds.}
\end{figure}

{Since the likely physical  origin of the disk effective viscosity
is  turbulence,   in  particular  the  turbulence   generated  by  the
magneto-rotational  instability (MRI)  \citep{bh91}, the  torque acting
upon the planet does not have a value constant in time as in a laminar
disk, but rather displays large temporal fluctuations arising from the
density  perturbations in  the  planet vicinity.  It  is important  to
assess  the impact  of these  fluctuations on  the  mechanism outlined
here. In  the bulk of  a disk with  a smooth surface  density profile,
these torque fluctuations can yield an erratic behavior of the planet
semi-major    axis   referred    to   as    ``stochastic   migration''
\citep{np2004,nelson2005}.   In   what  follows  we   use  the  torque
fluctuation  estimates  of  \citet{np2004} and  \citet{nelson2005}  in
order to determine  whether despite of its semi-major  axis jitter the
planet is confined to the trap vicinity, or whether it is unlocked the
from the  trap position.  We use  a very simple  prescription, that we
detail in  appendix~\ref{turbmc}, to perform  Monte-Carlo calculations
of   the   planet   semi-major    axis   over   a   period   of   time
$t=10^7\;\Omega^{-1}$.   We place  initially the  planet at  the outer
stable fixed point,  and we declare it trapped if  over that amount of
time it never  goes through the inner, unstable  fixed point, while we
declare it  not trapped if  it happens to  go through the  inner fixed
point.  While appendix~\ref{turbmc} presents the technical details of
our  calculations, we  discuss below  a number  of issues  relevant to
these calculations.
\begin{itemize}
\item Can the  torque acting upon the planet be  considered as the sum
of  the torque in  a laminar  disk and  the fluctuations  arising from
turbulence~? \citet{nelson2005} finds that this might not be the case,
at least  as long  as the differential  Lindblad torque  is concerned,
since  the  corotation region  is  not  sufficiently  resolved in  his
calculations; as an alternative explanation  to the lack of the torque
convergence  to  the laminar  value,  \citet{nelson2005} suggests  the
existence of very low frequency components of the torque fluctuations.
As this problem is essentially open (and totally unaddressed as far as
the corotation torque  is concerned), we chose to  describe the torque
acting upon  the planet by the sum  of the torque in  the laminar case
and of the torque fluctuations due to the turbulence.  
\item  The amplitude  ${\cal A}$  of  the torque  fluctuations can  be
estimated by considering  a density fluctuation of order  unity with a
length   scale  of   order   $H$,   at  a   distance   $H$  from   the
planet~\citep{np2004,nelson2005}  and   found  to  amount   to  ${\cal
A}=G\Sigma r$, where $G$ is  the gravity constant and $r$ the distance
to  the central  object.  This is  the  value that  we  adopt for  our
calculations.
\item The time scale of the fluctuations is of particular importance
to assess the torque convergence towards its laminar
value. \citet{nelson2005} finds in his calculations that the torque
fluctuations due to turbulence contain significant power at very low
frequency, and suggests that this could be linked to global
communication across the disk on the viscous timescale. Here our
situation is rather different, as the planet lies on a radially
localized structure, with a radial extent $\lambda$. Fluctuations with
frequency lower than $\sim \lambda/c_s$ would imply turbulent structures
larger than the trap width, and therefore would imply a radial
displacement of the trap itself along the lines of section~\ref{sec:trap}.
Here we choose a correlation timescale for the turbulence of 
$t_c=8\Omega^{-1} \sim 2\lambda/c_s$ for $\lambda\sim 4H$.
\item Both the tidal torque and the torque fluctuations scale with the
disk surface  density, so that the  time needed for  the time averaged
torque to  reach convergence does  not depend on the  latter. However,
the spread  of the probability  density of the planet  semi-major axis
after that timescale does depend  on the disk surface density.  If the
expectancy of  the distance of the  planet to its  initial location is
larger than  the trap  width, then the  planet can certainly  be lost,
while  if it  is  much smaller  than  the trap  width,  the planet  is
certainly trapped. Our study therefore amounts to a disk critical mass
search.   For a  given trap  profile, a  given planet  mass  and given
fluctuation characteristics  ${\cal A}\propto\Sigma$ and  $t_c$, there
exists a  critical disk mass so  that a lighter disk  keeps the planet
trapped, while a  heavier disk is unable to lock it  and the planet is
likely to go  through the inner unstable fixed  point. We perform this
search in a dichotomic manner until  we reach a precision of $5$~\% on
the disk  mass, and  we repeat the  calculation with  different random
seeds, so as to get an idea of the precision on the disk critical mass
estimate.  Our  results are displayed  on Fig.~\ref{fig:trapturb}. The
``disk   outer   mass''  is   defined   as  $M_{\rm   outer\;disk}=\pi
r_t^2\Sigma_{\rm out}$,  where $\Sigma_{\rm out}$ is  the disk surface
density on the  outside of the surface density  transition.  We see on
this  figure that  the deep  cavity with  $F=7$  unconditionally traps
$M=1$ to $15\;M_\oplus$ protoplanets in  a disk with a density profile
of the minimum  mass solar nebula, at any distance  up to $\sim 5$~AU,
while the shallow  surface density transition would be  able to retain
in  falling  bodies  only  in  the very  central  regions  ($r_t\simeq
0.01$~AU,  if  the nebula  extends  that far  in)  of  such a  nebula.
Further out  in the disk, at  $r=0.1$~AU, a shallow  transition in the
MMSN would only capture embryos with $M>4\;M_\oplus$, etc.
\end{itemize}
It is  not unreasonable to  assume that these simple  calculations are
conservative, for the following reasons:
\begin{itemize}
\item We have  assumed the torque fluctuations to  scale only with the
disk surface density, which implies  that all the disk material in the
trap vicinity is  turbulent. There is at least  one realization of the
trap, which  is the inner  edge of a  dead zone (detailed at  the next
section), for which this statement is wrong.  In our example, the trap
stable  fixed point  corresponds to  a location  with a  large surface
density,  therefore  located  rather  on  the  ``dead''  side  of  the
transition,  i.e. where  turbulent  transport of  angular momentum  is
inefficient and hence where the  turbulence should be much weaker than
contemplated in our simple Monte Carlo calculations.
\item We  recall that  our large potential  smoothing length  tends to
significantly  underestimate  the corotation  torque,  while the  trap
ability   to   lock  infilling   bodies,   for   a  given   turbulence
characteristics, strongly depends on  the corotation torque peak value
at the surface density transition.
\end{itemize}
We  conclude this section  with two  remarks.  Firstly,  the numerical
simulations   undertaken  by  \citet{np2004}   and  \citet{nelson2005}
neglect,  for  reasons   of  computational  speed,  the  gravitational
potential  $z$  dependency,  which  means  that their  disks  have  no
vertical  stratification. Future  numerical experiments  with vertical
stratification  may lead  to different  characteristics of  the torque
fluctuations.   Secondly,  other   sources  of  turbulence  have  been
suggested  in  protoplanetary disks,  such  as  the global  baroclinic
instability  \citep{klahr2004},   which  may  have   different  torque
fluctuations  statistics,  and  for  which  the  trap  efficiency  may
therefore  be different. To  date this  is essentially  an unaddressed
issue.}

\section{Discussion}
\label{sec:discuss}
The mechanism exhibited here is generic,  since it plays a role at any
positive disk surface density  transition, even shallow ones, with the
cautionary  remark that  some turbulence  is required  to  prevent the
corotation torque saturation, {and that a condition is required on
the  disk mass  to prevent  the  stochastic migration  induced by  the
turbulence to  overcome the trap  effect}.  Although our  knowledge of
the detailed properties of protoplanetary disks is still incipient, we
already  know a  number of  locations for  transitions  exhibiting the
above  properties,  which constitute  potential  traps for  in-falling
embryos:

\begin{itemize}
\item At the inner edge of a dead zone~\citep{gammie96,fromang02}, the
surface density  is expected  to be significantly  larger in  the dead
zone than in  the ionized inner disk, while  the inner disk turbulence
provides the  dissipation required to maintain  the corotation torque.
If  the  dead zone  inner  radius  changes  with time  (since  layered
accretion involving a dead zone is thought not to be a steady process)
the objects trapped  at the fixed point follow  the transition radius,
as  shown in  section~\ref{sec:trap}.   The edges  of  dead zones  are
thought,  in the  totally  different context  of  runaway or  type~III
migration  \citep{mp03}, to  be  able  to halt  or  reverse a  runaway
episode  \citep{arty04}. Here, in  our context  of small  mass, type~I
migrating objects, we  argue that the inner edge of  a dead zone halts
type~I migrating bodies and efficiently traps them.
\item Sizable  surface density  transitions can also  be found  at the
transition between  an inner disk  region where jets are  launched and
the  outer  viscous  disk~\citep{ferreira95}.  Indeed,  the  accretion
velocity in the  jet emitting disk (hereafter JED)  region is entirely
due  to the  jet torque  and becomes  much larger  than in  a standard
accretion disk (hereafter SAD). Mass flux conservation then leads to a
surface density jump at the  transition radius between the SAD and the
JED.  This  radius  has been  inferred  to  vary  from  0.3 to  a  few
astronomical units  in some  T Tauri stars\footnote{In  fact, velocity
shifts detected in jets are  interpreted as being a rotation signature
which then allows one to derive  the outer jet launching radius in the
underlying accretion  disk. This radius corresponds  to our transition
radius.}~\citep{coffey04,pesenti04}.  As  all  known embedded  objects
have  jets, the  existence  of  such a  transition  radius provides  a
natural trap for in-falling embryos over the jet lifetime.
\item  We  finally  mention the  outer  edge  of  a  gap opened  by  a
preexisting giant  protoplanet. The  latter can trap  planetesimals at
mean motion  resonances \citep{hw96}, but  also at the  large positive
surface density gradient found on its gap outer edge.
\end{itemize}

We  note that  the first  two possibilities  might not  coexist  in an
accretion disk.   Not all T Tauri  stars show indeed  evidence of high
velocity jets~\citep{hartigan95}.  For  these jet-less stars, the whole
disk is  likely to resemble  a SAD with  a dead zone in  its innermost
region.   The protoplanet  trap would  then be  realized at  the inner
radius of  the dead  zone, namely only  if there  is a SAD  settled at
smaller radii.   Objects with a dead  zone extending down  to the star
would  have no  trap.  On  the  contrary, young  stars producing  jets
require the presence of a JED  in the inner disk region. In that case,
the trap is located at the outer JED radius (the JED itself may extend
down to the star with no impact  on the trap).  Note also that since a
JED is less  dense than a SAD,  it is unlikely that a  dead zone would
establish (Ferreira et al. in prep.).

The likely  existence of such planetary traps  in protoplanetary disks
has important implications for  the formation of giant planets. First,
it constitutes a workaround to  the timescale conflict between a solid
core build up (which is slow)  and its migration to the central object
(which  would   be  fast,  if  the   object  was  not   stopped  at  a
trap). Second,  a large amount  of solid embryos should  accumulate at
the trap  location. This should  speed up the  build up of  a critical
mass core  ($M_p \simeq 15\;M_\oplus$) which could  accrete the nebula
gas and convert itself into a giant planet. The dynamics of the set of
embryos  trapped at  the  jump is  however  beyond the  scope of  this
communication, and will be presented in a forthcoming work.

We have seen in  section~\ref{sec:analytic} that the corotation torque
value is  boosted not only by  the large surface  density gradient but
also by the  alteration of the rotation profile,  which yields a large
gradient of $B$, with both  large minimal and maximal values since the
gradient  of  $B$  involves  the  second  derivative  of  the  angular
velocity.  One  may wonder  whether at a  disk opacity  transition, in
which there is  a disk temperature jump and  hence a pressure gradient
radial jump, the  alteration of the rotation profile  is sufficient to
yield  a peak  value of  the  corotation torque  that counteracts  the
Lindblad torque.  We have  undertaken subsidiary calculations in which
we have a power law surface  density profile and a sound speed profile
with a  localized radial  jump. We found  that we  need a jump  of the
aspect ratio of $\sim 1.5$ over  $4H$ in order to halt migration. This
ratio is too large and is  unrealistic in a SAD. Indeed, in a standard
accretion disk in which  energy is transported vertically by radiative
diffusion,  the  central  temperature  scales as  $\tau^{1/8}$,  where
$\tau$ is the optical depth. As a consequence, even in the vicinity of
an  opacity  transition, the  temperature  profile  exhibits a  smooth
behavior  and the  corotation  torque is  not  sufficiently strong  to
counterbalance the Lindblad  torque. Nevertheless, \cite{menou04} have
found that the opacity  transitions may induce a significant reduction
of the differential Lindblad torque, but they explicitly neglected the
corotation  torque  in  their  study.   It would  be  of  interest  to
investigate whether the corotation torque boost at these locations can
be sufficient to halt migration.

Remarkably, the scale  height of a JED is always  smaller than that of
the  SAD, so  that the  transition between  these two  flows naturally
yields a jump  in the disk aspect ratio.  This is  due to two effects.
First, a JED feeds its jets  with most of the released accretion power
so  that only a  small fraction  of it  is radiated  away at  the disk
surface:  the disk  is cooler.   Second, in  addition to  gravity, the
large   scale    magnetic   field   vertically    pinches   the   disk
\citep{ferreira97}.

We also stress that at the inner dust puffed-up rims of protoplanetary
disks~\citep{isella05}, there  can be a  large radial gradient  of the
gas temperature, which may be sufficient to trap migrating planets.

We finally emphasize that the  mechanism presented here does not allow
low mass objects to proceed inside  of a cavity, but rather traps them
on the cavity edge. This is quite different from what has been studied
in the  context of gap  opening planets, which migrate  further inside
until   they    reach   the   $2:1$   resonance    with   the   cavity
edge~\citep{KL02}.

\section{Conclusions}

We have shown that the  type I migration of $M_p<\;15M_\oplus$ planets
can be  halted at radially  localized disk surface density  jumps (the
surface density  being larger on the  outside of the  jump). {Some
turbulence  is   then  required  to  prevent   the  corotation  torque
saturation.  The  latter tends to  induce stochastic migration  of the
embryos, and therefore tends to  act against the trap. The lighter the
disk, the  easier it is  to keep an  embryo trapped, since  its radial
jitter over the timescale needed  to reach torque convergence is small
and  can  be  confined  to  the  vicinity of  the  trap  stable  fixed
point. Inversely, the lighter the  embryo, the more difficult it is to
trap it:  in the zero mass limit,  the embryo has no  tidal torque and
only feels  the torque fluctuations  arising from turbulence.  We have
shown that a jump with a large surface density ratio efficiently traps
embryos above one Earth mass up even in relatively massive disks, such
as  the MMSN  at  $r=5$~AU,  while shallow  jumps  (a $40$~\%  surface
density increase over $\sim 4$ disk scale heights) will retain embryos
of a few Earth  masses and more only in light disks,  such as the MMSN
at $r=0.1$~AU.}  We have illustrated by means of numerical simulations
that  a surface  density jump  traps in-falling  embryos, and  that it
drives them  along with it if  it moves radially as  the disk evolves.
Such a jump can occur at different locations in a protoplanetary disk:
it can be found at  the transition between the standard accretion disk
solution (SAD)  and the jet  emitting disk (JED),  thought to be  at a
distance $0.3$  to $1-2$~AU  from the central  object. It can  also be
found at the inner edge of a dead zone, where one expects a very large
ratio of  outer to inner surface  densities.  It may also  be found at
the outer  edge of  a protoplanetary gap  if one giant  planet already
exists in the  disk. We have stressed that  these planetary traps have
two important consequences: (i) they  can hold the type~I migration of
protogiant cores  over the timescale  needed to overcome  the critical
mass ($\sim 15\;M_\oplus$) over  which rapid gas accretion begins onto
these cores, which are then  quickly converted into giant planets that
should enter the slow, type~II migration regime; (ii) as solid embryos
accumulate  at the  trap, the  build up  of critical  cores  should be
faster than elsewhere in the disk. Finally, we mentioned that although
temperature radial  jumps in principle  also yield large peaks  of the
corotation  torque, they  are unlikely  to be  able to  counteract the
Lindblad torque  by themselves, as the  temperature gradients required
for that are unrealistically large in a standard accretion disk.

\appendix

\section{Prescription for the 1D calculation of the planet semi-major
axis time evolution in the presence of torque fluctuations}
\label{turbmc}
The planet semi-major axis is evolved in time using an Euler method:
\begin{equation}
a(t+\Delta t)=a(t)+\Delta t\frac{\Gamma_l+\Gamma_f}{2B[a(t)]a(t)},
\end{equation}
where $a(t)$ is the planet  semi-major axis at date~$t$, $\Gamma_l$ is
the torque  as measured  in a  laminar disk for  a planet  in circular
orbit of radius $a$,  linearly interpolated from the results displayed
either   in  fig.~\ref{fig:tqnumcavit}a   or  \ref{fig:cavitshallow}a,
$\Gamma_f$ is  the torque fluctuation  due to turbulence, and  where a
Keplerian profile is assumed for $B$  (this is grossly wrong as far as
its radial  derivative is concerned, as  we saw in the  main text, but
this  is  largely  sufficient   for  our  purpose  here).  The  torque
fluctuation is evaluated as follows:
\begin{equation}
\Gamma_f = G\Sigma_m[a(t)]a(t)[2V-1],
\end{equation}
where $\Sigma_m[a]=\max[\Sigma(r)]_{r\in[a-H,a+H]}$,  and where $V$ is
a  random variable uniformly  distributed over  $[0,1]$. The  value of
$\Gamma_f$ is kept constant over  $t_c$. We use the above prescription
for $\Sigma_m[a]$  since the planet is  located in a  region where the
surface density gradient is large. The largest torque fluctuations may
arise  from regions  located within  $a-H$ to  $a+H$ from  the central
object, and  where the surface density may  be significantly different
from the  one at the planet  location. Each realization  of the random
variable is independent  of the previous ones. We  anticipate that the
results should  not depend  on the torque  fluctuation law,  since the
Central  Limit Theorem  states that  the cumulative  torque  law tends
towards a normal law \citep{nelson2005}.

% 
% References
%

%
\end{document}